\documentclass{article}
\usepackage{spconf,amsmath,graphicx}

\usepackage{amsthm,amssymb}
\usepackage{bm}
\usepackage{color}
\usepackage{xcolor}
\usepackage{multirow}
\usepackage{float}
\usepackage{booktabs}
\usepackage{cite}
\usepackage{verbatim}
\usepackage{hyperref}
\hypersetup{
  colorlinks=true,    
  linkcolor=black,     
  citecolor=black,    
  urlcolor=black 
}
\usepackage{cleveref}

\title{Enhancing Audio Generation Diversity with Visual Information}
\name{Zeyu Xie,
        Baihan Li,
      Xuenan Xu,    
       Mengyue Wu$\dag$, Kai Yu$\dag$\thanks{$\dag$Mengyue Wu and Kai Yu are the corresponding authors. This work has been supported by National Natural Science Foundation of China (No.92048205), the Key Research and Development Program of Jiangsu Province, China (No.BE2022059), and Shanghai Municipal Science and Technology Major Project (2021SHZDZX0102). }}
\address{
MoE Key Lab of Artificial Intelligence\\
X-LANCE Lab, Department of Computer Science and Engineering\\
AI Institute, Shanghai Jiao Tong University, Shanghai, China
}

\begin{document}
%
\maketitle
\begin{abstract}
Audio and sound generation has garnered significant attention in recent years, with a primary focus on improving the quality of generated audios. 
However, there has been limited research on enhancing the diversity of generated audio, particularly when it comes to audio generation within specific categories. 
Current models tend to produce homogeneous audio samples within a category. 
This work aims to address this limitation by improving the diversity of generated audio with visual information.
We propose a clustering-based method, leveraging visual information to guide the model in generating distinct audio content within each category. 
Results on seven categories indicate that extra visual input can largely enhance audio generation diversity.
Audio samples are available at \href{https://zeyuxie29.github.io/DiverseAudioGeneration/}{\textcolor{cyan}{\textit{DemoWeb}}}.
\end{abstract}
\section{Introduction}
\label{sec:intro}

Recently, generative tasks have witnessed significant advancements, with Text To Audio generation (TTA) tasks emerging as one of the prominent topics. 
TTA aims at generating corresponding audio content based on textual conditioning, which can either be category labels or free-form natural language descriptions.
TTA exhibits extensive applications: data augmentation~\cite{choi2023foley}, providing sound effects for movies and games~\cite{kong2019acoustic} as well as customized generation for diverse needs.
The generation performance is influenced by variations in conditions, thus posing a challenge in producing high-quality and diverse audio from limited input information.


\begin{figure}[htbp]
  \centering
  \centerline{\includegraphics[width=0.85\linewidth]{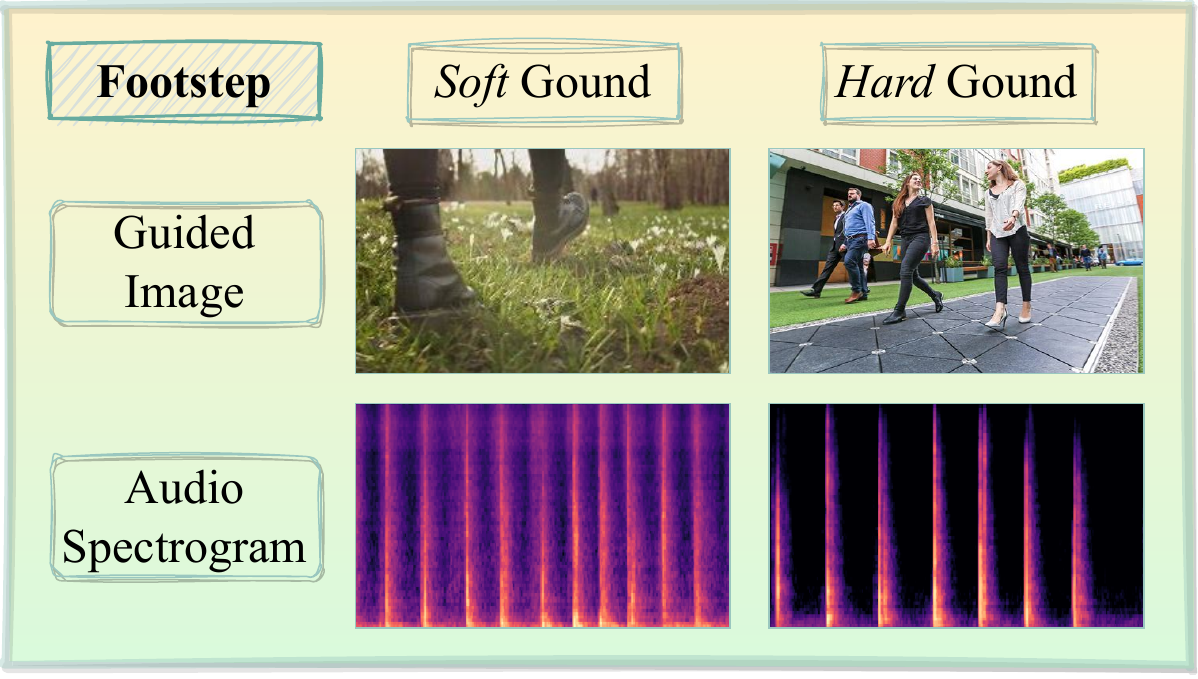}}
  \caption{Vision-guided generated samples.
}
  \label{fig:intro_example}
\end{figure}

DCASE2023 task7~\cite{choi2023foley} provides a dataset for \textbf{category-based audio generation}, where the textual conditions are sound event labels.
However, the generative model trained under this dataset exhibited a proclivity to generate audios with a uniform style, which corresponds to the predominant style within each category.
Nevertheless, audio samples in the training set are much more diverse than model-generated ones.
Two possible factors might give rise to the homogeneous generation patterns: 1) the great diversity in training data is difficult to implicitly model given limited training samples; 2) the input single category label cannot encompass the various types of information within a category.
For instance, the sound of footsteps on a \textit{hard} surface differs in timbre from that on a \textit{soft} surface.
Such diversity is difficult to learn and generate under the one-to-many learning paradigm.
\begin{figure*}[htbp]
  \centering
  \centerline{\includegraphics[width=0.92\textwidth]{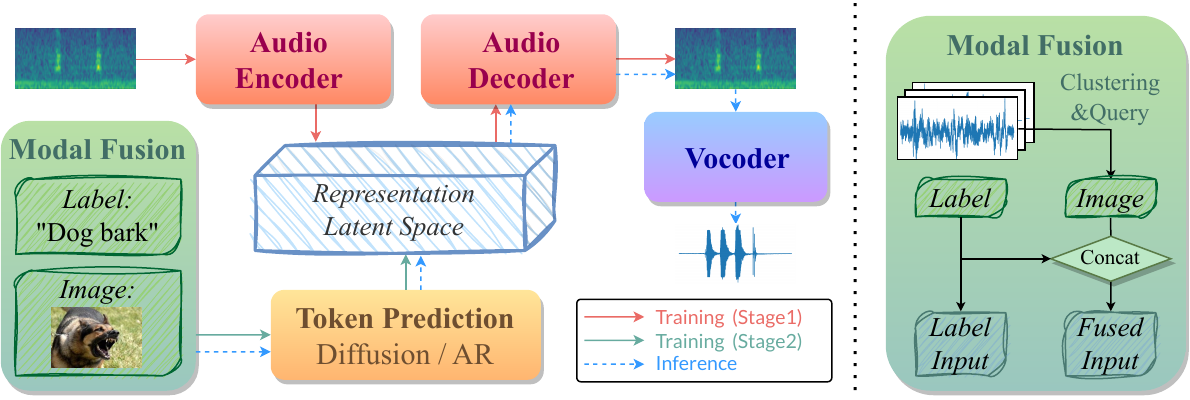}}
  \caption{The diagram of audio generation framework.
  The input condition is processed via a modal fusion module.
  Red and green arrows represent the model training, including {\textcolor[RGB]{234,107,102}{1) red:}} VAE/VQ-VAE learning for audio representation, and {\textcolor[RGB]{103,171,159}{2) green:}} token prediction based on input conditions.
  {\textcolor[RGB]{51,153,255}{Blue}} arrows show the inference process: the token prediction model predicts the audio representation, which is then restored to a spectrogram by the VAE/VQ-VAE decoder. 
  Finally, the audio is generated by a vocoder.
}
  \label{fig:results}
\end{figure*}

To better measure the within-category audio diversity in the training set, we first employ an acoustic-based \textbf{unsupervised clustering} method to establish a more fine-grained categorization, partially mitigating the one-to-many problem. 
On the other hand, the natural audio-visual alignment inspires \textbf{additional visual input} to enhance audio diversity. 
For instance, the timbre difference between footsteps on \textit{soft} and \textit{hard} surfaces can be better differentiated given corresponding images.
As it involves the assessment of different ground materials and types of footwear, which are less conveniently described in free-text form.
In light of this, we further query images from the internet and assign them to each fine-grained subcategory, which are fed alongside the category labels to the model. 

We propose a framework that innovatively integrates visual information into category-to-audio generation tasks via a clustering-based approach.
The experimental results demonstrate that, given more fine-grained and richer visual input, the generated audio \textbf{diversity has increased}, further elevating the overall quality of the generated audio clips.
Such a vision-guided method \textbf{is adaptable to} various mainstream generative models, enhancing diversity in the generated audios. 

\section{Vision-guided audio generation}
\label{sec:method}

As \Cref{fig:results} shows, the generation framework consists of 1) a modal fusion module for integrating visual information; 2) audio representation models to compress audio and learn latent representations; 3) token prediction models to predict the compressed representation from input conditions.

During inference, the latent audio representation predicted by token prediction models is reconstructed into a spectrogram by the decoder of audio representation models, which is further transformed to waveform via a pre-trained vocoder.


\subsection{Modal Fusion}
\label{ssec:input_process}

The input category label $\mathcal{S}_{i}$ is encoded into an embedding $\mathcal{I}_{label}$ via a lookup table.
However, within a single category, sounds may exhibit similar features but distinct details.
For instance, the barking of mature dogs and puppies differ.
To better characterize different patterns for diverse generation, 
spectral clustering~\cite{wang2023wespeaker} is implemented in each category, based on audio features (e.g. mel spectrum, spectrum centroid, spectrum bandwidth, zero crossing rate, root mean square, spectrum flatness and mel frequency cepstral coefficient).
Each category is clustered into $2$ to $4$ fine-grained subcategories, with the number of subcategories automatically determined through spectral clustering.

The natural audio-visual alignment and the richer details present in the visual modality inspire the inclusion of additional visual input to enhance audio diversity.
For each subcategory $\mathcal{S}_{i,k}$, relevant images are retrieved manually from online sources and allocated, resulting in $2$ to $24$ images for each subcategory.
The image retrieval process can be automatically completed by retrieval models~\cite{wilkins2023bridging} when scaling to large datasets.
The image information is integrated using two approaches: averaging or manually selecting one prototype image.
CLIP~\cite{radford2021learning} is utilized to extract subclass image features, which are concatenated with parent class label $\mathcal{I}_{label}$ to create a fused visual-textual input: 
\begin{equation}
    \begin{split} 
  \label{eqn:input}
    \mathcal{I}_{avg} = \mbox{concat}(\mathcal{I}_{label}, \frac{1}{|\mathcal{S}_{i,k}|}\sum_{v \in \mathcal{S}_{i,k}}^{}{\mbox{CLIP}(v)})  \\
    \mathcal{I}_{pro} = \mbox{concat}(\mathcal{I}_{label}, {\mbox{CLIP}(v_{prototype})}) 
    \end{split} 
\end{equation}


\subsection{Audio Representation}
\label{ssec:representation }

In audio generation tasks, it is challenging to generate spectrograms directly.
One solution is to represent the spectrogram in latent space via Variational Autoencoders (VAE) or Vector Quantized VAE (VQ-VAE) to reduce the burden on the generation model~\cite{yang2023diffsound,liu2023audioldm,kreuk2022audiogen,huang2023make,liu2021conditional}. 

VAE encoder compresses the spectrogram $ {\mathcal{A}}\in \mathbb{R}^{M \times T}$ into a latent representation $ {\mathcal{P}}\in \mathbb{R}^{M/2^c \times T/2^c \times D}$, 
where $M, T, D, c$ denotes the number of mel bands, the sequence length, the codebook dimension, and the compression ratio, respectively. 
The latent representation $ {\mathcal{P}}$ is partitioned into $ {\mathcal{P}}_\mu$ and $ {\mathcal{P}}_\sigma \in \mathbb{R}^{M/2^c \times T/2^c \times D/2}$, representing the mean and variance of the VAE latent space. 
The decoder reconstructs the spectrogram $\hat{ {\mathcal{A}}}$ based on samples drawn from the distribution $\hat{ {\mathcal{P}}}= {\mathcal{P}}_\mu+ {\mathcal{P}}_\sigma \cdot \mathcal{N}(0,1)$.

Compared with VAE, VQ-VAE uses a codebook to quantize $\mathcal{P}_{m,t}$ into codebook entries, which is utilized to reconstruct the spectrogram by the decoder.





\subsection{Token Prediction}

Token prediction model learns to predict representations $\hat{ {\mathcal{P}}}$ based on conditions $ {\mathcal{I}}$.
Auto-regressive models~\cite{liu2021conditional, kong2019acoustic, kreuk2022audiogen} and diffusion~\cite{yang2023diffsound,liu2023audioldm,huang2023make, ghosal2023text, choi2023hyu} are commonly utilized. 

\textbf{Auto-regressive Model}
Transformer~\cite{vaswani2017attention} is employed due to its efficiency and robust modeling capabilities in sequence processing tasks.
Transformer predicts the distribution of tokens based on previous ones and the feature sequence.
Self-attention and cross-attention are employed to encode previous tokens into embeddings and incorporate information between the input feature and embeddings, respectively.

\textbf{Latent Diffusion Model} LDM has demonstrated its capacity to produce high-quality audio in TTA tasks~\cite{liu2023audioldm,huang2023make, ghosal2023text, choi2023hyu}.
LDM encompasses a forward process that transforms the latent distribution into a Gaussian distribution through noise injection, followed by a reverse diffusion process responsible for denoising and spectrogram generation.
The transition probabilities of the Markov chain in the forward process with noise schedule $\{\beta_n:0 < \beta_n<\beta_{n+1}<1\}$ are:

\begin{align}
    q(\mathcal{P}_n|\mathcal{P}_{n-1})\triangleq \mathcal{N}(\sqrt{1-\beta_n}\mathcal{P}_{n-1},\beta_n \mathbf{I})\\
    \mathcal{P}_n=\sqrt{\bar{\alpha}_n}\mathcal{P}_0 + \sqrt{1-\bar{\alpha}_n}\epsilon_n
\end{align}


where $\alpha_n = 1 - \beta_n, \bar{\alpha}_n=\prod_{i=1}^{b}{\alpha_i}$, $\epsilon_n$ is sampled from the normal distribution $\epsilon \sim \mathcal{N}(0,1)$.
At final time step $N$, $\mathcal{P}_N $ follows an isotropic Gaussian noise.
With input $\mathcal{I}$ and a weight $\lambda_n$ related to Signal-to-Noise Ratio (SNR)~\cite{hang2023efficient}, the reverse process learns to estimate noise:
\begin{equation}
  \label{eqn:diffusion_r1}
        \mathcal{L}=\sum_{n=1}^{N}{\lambda_n \mathbb{E}_{\epsilon_n, \mathcal{P}_0}||\epsilon_n - \epsilon_{\theta}(\mathcal{P}_n, \mathcal{I}) ||}
\end{equation}
where $\epsilon_{\theta}$ denotes the noise estimation network, which can be employed to reconstruct $\mathcal{P}_0$ from $\mathcal{P}_N \sim \mathcal{N}(0,1)$.



\begin{table*}[t]
    \centering
    \caption{Diversity: Mean squared distance (MSD$_f$) $\uparrow$. ``+" signifies that visual information is integrated. }
    \begin{tabular}{cc|ccccccc|c}
    \toprule
    \multirow{2}*{ID}&\multirow{2}*{System} & \multirow{2}*{dog bark} &\multirow{2}*{footstep}& \multirow{2}*{gunshot} & \multirow{2}*{keyboard} & moving motor & \multirow{2}*{rain} & \multirow{2}*{sneeze cough} & average\\
    & & & & &  & vehicle & & & MSD$_{f}$\\
    \midrule
   A0 & \textbf{Vae-Diff-g2} & $\mathbf{31.05}$ & $6.00$ & $9.12$ & $5.61$ & $21.54$ & $10.69$ & $13.16$ & $13.88$\\
   A1 & $+$average & $21.27$ & $\mathbf{11.11}$ & $29.58$ & $6.20$ & $32.70$ & $12.76$ & $16.80$ & $18.63$\\
   A2 & $+$prototype & $19.74$ & $10.75$ & $\mathbf{43.94}$ & $\mathbf{8.75}$ & $\mathbf{44.49}$ & $\mathbf{15.33}$ & $\mathbf{18.81}$ & $\mathbf{23.12}$\\
   \midrule
   B0 & \textbf{Vae-Diff-g3} & $\mathbf{39.49}$ & $3.73$ & $9.94$ & $5.36$ & $35.28$ & $10.11$ & $25.97$ & $18.55$\\
   B1 & $+$average & $27.83$ & $11.18$ & $55.76$ & $6.21$ & $82.56$ & $13.83$ & $40.43$ & $33.97$\\
   B2 & $+$prototype & $26.83$&$\mathbf{11.36}$&$\mathbf{75.26}$& $\mathbf{9.46}$&$\mathbf{94.89}$&$\mathbf{17.00}$&$\mathbf{44.36}$ & $\mathbf{39.88}$\\
   \midrule
   C0&\textbf{Vqvae-Trm} & $26.77$ & $12.73$ & $32.97$ & $12.01$ & $18.60$ & $12.05$ & $14.58$ & $18.53 $ \\
   C1&$+$average  & $27.67$ & $\mathbf{19.83}$ & $36.39$ & $10.50$ & $22.07$ & $15.35$ & $22.14$ & $21.99$\\
   C2&$+$prototype  &$\mathbf{30.61}$ & $15.89$ & $\mathbf{39.37}$ & $\mathbf{14.44}$ & $\mathbf{24.30}$ & $\mathbf{15.96}$ & $\mathbf{25.78}$ & $\mathbf{23.76}$\\
    \bottomrule
    \end{tabular}
    \label{tab:msd}
\end{table*}

\begin{table}[t]
    \centering
    \caption{Objective evaluation}
    \label{tab:all}
    \begin{tabular}{c|cc|cc}
    
    \toprule
    \multirow{2}*{ID} & \multicolumn{2}{c}{Quality $\downarrow$} & \multicolumn{2}{|c}{Diversity $\uparrow$}\\
  \cmidrule(lr){2-3}\cmidrule(lr){4-5}
   &  Dev$_{FAD}$ & Test$_{FAD}$& MSD$_{f}$ & MSD$_{B}$\\
   \midrule
   
   A0 &  $\mathbf{4.81}$& $9.87$ & $13.88$ & $11.44$ \\
   A1 &   $5.08$ &$\mathbf{9.06}$ & $18.62$ & $11.88$ \\
   A2 &   $4.91$ & $9.12$ & $\mathbf{23.12}$ & $\mathbf{12.12}$ \\
   \midrule
   B0 &   $4.79$ & $11.16$ & $18.55$& $11.08$ \\
   B1 &  $4.76$&$\mathbf{10.23}$ & $33.97$ &$11.71$\\
   B2&$\mathbf{4.55}$&$10.76$&$\mathbf{39.88}$&$\mathbf{12.00}$\\
   \midrule
   C0 &  $5.73$ & $7.98$ & $18.53$& $15.24$ \\
   C1 &  $\mathbf{5.56}$&$\mathbf{7.55}$ & $21.99$ &$14.98$\\
   C2 &$6.14$&$8.05$&$\mathbf{23.76}$&$\mathbf{15.55}$\\
    \bottomrule
    \end{tabular}
\end{table}

\section{Experiment Setup}
\label{sec:exp}


\subsection{Generation Framework}
We use the dataset in DCASE2023 task 7.
Two frameworks are implemented for prediction.
The vocoder is a pre-trained HiFi-GAN~\cite{kong2020hifi}.

\textbf{VAE \& LDM} We adopt the pre-trained VAE model from Liu et al.~\cite{liu2023audioldm}.
The LDM has a similar structure with Ghosal D et al.~\cite{ghosal2023text} but employs fewer parameters, with attention dimension $\{4, 8, 16, 16\}$ and block channels $\{128, 256, 512, 512\}$. 
AdamW optimizer is used to train the LDM for 80 epochs.
The learning rate is set to $3 \times 10^{-5}$ with a linear decay scheduler.
Classifier-free guidance~\cite{ho2022classifier,nichol2021glide,liu2023audioldm} is adopted for controllable generation.
The guidance scale is set to $2$ or $3$, corresponding to \textit{Vae-Diff-g2} and \textit{Vae-Diff-g3} in \Cref{tab:msd}.

\textbf{VQ-VAE \& Transformer} The VQ-VAE model follows DCASE baseline~\cite{liu2021conditional}, so is the loss function.
The Transformer comprises one encoder layer and two decoder layers. 
Cross-entropy loss is adopted.
The hidden dimension, feed-forward dimension and attention heads are set to $512$, $2048$ and $8$ respectively.
Both VQ-VAE and Transformer are trained in 800 epochs using Adam optimizer with a learning rate $3 \times 10^{-4}$.
The model with the minimum loss is utilized for generation and denoted as \textit{Vqvae-Trm}.

\subsection{Evaluation Metrics}
Objective and subjective metrics are employed to evaluate the quality and diversity of the generated audios, facilitating a comprehensive evaluation of incorporating visual modality. 

\textbf{Objective-Quality}
Fr$\acute{\text{e}}$chet Audio Distance (FAD)~\cite{kilgour2019frechet} is commonly used in TTA tasks, which is based on the distribution of generated and reference samples.
The DCASE challenge offers the distribution of the unreleased test set for evaluation, denoted as Test$_{FAD}$.
A common evaluation protocol in TTA is that the closer the generated distribution is to the training distribution, the better its quality.
Thus Dev$_{FAD}$ is further computed between the training data and the generated audios to comprehensively assess the models.


\textbf{Objective-Diversity} The diversity among audio samples can be quantified using the Mean Squared Distance (MSD) metric, which measures the pairwise distance distribution of generated audios.
The audio features, consistent with~\Cref{ssec:input_process}, are used to compute MSD$_{f}$.
The self-supervised BEATs~\cite{chen2022beats} model, which extracts non-category-specific audio features, is employed to calculate MSD$_{B}$.

\textbf{Subjective}
Mean Opinion Score (MOS) is conducted from three perspectives: event accuracy, generation diversity and audio naturalness. 
For each sound category, evaluators assess a set of $3$ generated audio samples. 
The results are derived as the mean scores over $10$ evaluators.

\section{Result}
\label{sec:result}
The results are presented in Tables~\ref{tab:msd}, ~\ref{tab:all} and~\ref{tab:mos}. 
\Cref{tab:msd} elaborates the MSD$_f$ values for each category.
\Cref{tab:all} presents objective evaluation for overall generation quality and diversity, and \Cref{tab:mos} showcases the overall subjective MOS results.

\subsection{Diversity Enhancement}

Both objective and subjective metrics in Tables~\ref{tab:all} and~\ref{tab:mos} demonstrate an enhancement in the diversity of generated audios. 
Notably, the improvement is more significant when leveraging prototype image (A2, B2, C2) as opposed to averaged features (A1, B1, C1).
This underscores the effectiveness of incorporating visual information: when images are more representative, generative models capture finer details of subcategories, resulting in more diverse audio generation.
It can be further inferred that employing different visual information enables controllability in audio generation.

Delving into the details, \Cref{tab:msd} reveals a significant improvement across the majority of categories, particularly evident in categories with originally lower MSD values such as ``footstep" and ``keyboard".
This bears substantial significance, as the introduction of visual information supplements the original model's shortcomings in generating diversity.
For instance, the baseline model, which relies solely on labels as input, predominantly generates common keyboard sounds in the ``keyboard" category. 
In contrast, the proposed method can generate distinguishable sounds for mechanical and office keyboards by incorporating additional visual information.

\subsection{Overall Quality}
The quality metrics in \Cref{tab:all}, as well as the accuracy and naturalness metrics in \Cref{tab:mos}, indicate that the quality of the audio generated with visual guidance is comparable to that generated solely using category labels.

It is noteworthy that diffusion-based models inherently provide a method to enhance diversity by employing different guidance scales. 
However, this comes at the cost of compromising the quality of audio generation.
As shown in \Cref{tab:all}, concerning the performance of \textit{Vae-Diff-g2} (A0) and \textit{Vae-Diff-g3} (B0), when the guidance scale increases, although the diversity improves, the quality declines. 
In contrast, with the incorporation of visual information, \textit{Vae-Diff-g2} (A1, A2) demonstrates diversity generation performance comparable to that of \textit{Vae-Diff-g3} (B0) while maintaining audio quality.

\begin{table}[t]
\centering
\caption{Subjective evaluation, tested on the \textit{Vae-Diff-g3} as well as the models incorporating visual information.}
\label{tab:mos}
    \begin{tabular}{cc|ccc}
    \toprule
    \multirow{2}*{ID} &\multirow{2}*{System} &\multicolumn{3}{c}{MOS metrics $\uparrow$} \\
    \cmidrule(lr){3-5}
    & & Accuracy &Diversity &Naturalness\\
    \midrule
    B0 &\textbf{Vae-Diff-g3} &  $\mathbf{4.37}$   & $3.28$  & $\mathbf{3.59}$\\
    B1 &$+$average &   $4.12$           &  $4.03$           & $3.11$\\ 
    B2 &$+$prototype &   $4.12$           & $\mathbf{4.24}$   & $3.49$\\
    \bottomrule
    \end{tabular}
\end{table}

\section{Conclusion}
\label{sec:conclusion}
This paper aims to enhance the diversity of category-based audio generation using a clustering-based approach that incorporates visual information.
When audio samples within a category exhibit significant distinctions, solely relying on category inputs can make it challenging for the model to establish one-to-many mapping relationships. 
Therefore, we employ clustering techniques to recognize distinct subgroups and align visual information as guidance.
Visual information is fused with the category label to help the model capture patterns among different sub-classes and generate more diverse audios. 
Experiments are conducted under two mainstream generative frameworks.
Evaluation results demonstrate that our approach, leveraging visual information, exhibits the capability to produce more diverse sound effects while preserving their quality.
Additionally, more representative images control the generation of more diverse audio.

\bibliographystyle{IEEEbib}
\bibliography{refs}

\begin{thebibliography}{10}

\bibitem{choi2023foley}
Keunwoo Choi, Jaekwon Im, Laurie Heller, Brian McFee, Keisuke Imoto, Yuki Okamoto, Mathieu Lagrange, and Shinosuke Takamichi,
\newblock ``Foley sound synthesis at the dcase 2023 challenge,''
\newblock {\em arXiv preprint arXiv:2304.12521}, 2023.

\bibitem{kong2019acoustic}
Qiuqiang Kong, Yong Xu, Turab Iqbal, Yin Cao, Wenwu Wang, and Mark~D Plumbley,
\newblock ``Acoustic scene generation with conditional samplernn,''
\newblock in {\em ICASSP 2019-2019 IEEE International Conference on Acoustics, Speech and Signal Processing (ICASSP)}. IEEE, 2019, pp. 925--929.

\bibitem{wang2023wespeaker}
Hongji Wang, Chengdong Liang, Shuai Wang, Zhengyang Chen, Binbin Zhang, Xu~Xiang, Yanlei Deng, and Yanmin Qian,
\newblock ``Wespeaker: A research and production oriented speaker embedding learning toolkit,''
\newblock in {\em ICASSP 2023-2023 IEEE International Conference on Acoustics, Speech and Signal Processing (ICASSP)}. IEEE, 2023, pp. 1--5.

\bibitem{wilkins2023bridging}
Julia Wilkins, Justin Salamon, Magdalena Fuentes, Juan~Pablo Bello, and Oriol Nieto,
\newblock ``Bridging high-quality audio and video via language for sound effects retrieval from visual queries,''
\newblock {\em arXiv preprint arXiv:2308.09089}, 2023.

\bibitem{radford2021learning}
Alec Radford, Jong~Wook Kim, Chris Hallacy, Aditya Ramesh, Gabriel Goh, Sandhini Agarwal, Girish Sastry, Amanda Askell, Pamela Mishkin, Jack Clark, et~al.,
\newblock ``Learning transferable visual models from natural language supervision,''
\newblock in {\em International conference on machine learning}. PMLR, 2021, pp. 8748--8763.

\bibitem{yang2023diffsound}
Dongchao Yang, Jianwei Yu, Helin Wang, Wen Wang, Chao Weng, Yuexian Zou, and Dong Yu,
\newblock ``Diffsound: Discrete diffusion model for text-to-sound generation,''
\newblock {\em IEEE/ACM Transactions on Audio, Speech, and Language Processing}, 2023.

\bibitem{liu2023audioldm}
Haohe Liu, Zehua Chen, Yi~Yuan, Xinhao Mei, Xubo Liu, Danilo Mandic, Wenwu Wang, and Mark~D Plumbley,
\newblock ``Audioldm: Text-to-audio generation with latent diffusion models,''
\newblock {\em arXiv preprint arXiv:2301.12503}, 2023.

\bibitem{kreuk2022audiogen}
Felix Kreuk, Gabriel Synnaeve, Adam Polyak, Uriel Singer, Alexandre D{\'e}fossez, Jade Copet, Devi Parikh, Yaniv Taigman, and Yossi Adi,
\newblock ``Audiogen: Textually guided audio generation,''
\newblock {\em arXiv preprint arXiv:2209.15352}, 2022.

\bibitem{huang2023make}
Rongjie Huang, Jiawei Huang, Dongchao Yang, Yi~Ren, Luping Liu, Mingze Li, Zhenhui Ye, Jinglin Liu, Xiang Yin, and Zhou Zhao,
\newblock ``Make-an-audio: Text-to-audio generation with prompt-enhanced diffusion models,''
\newblock {\em arXiv preprint arXiv:2301.12661}, 2023.

\bibitem{liu2021conditional}
Xubo Liu, Turab Iqbal, Jinzheng Zhao, Qiushi Huang, Mark~D Plumbley, and Wenwu Wang,
\newblock ``Conditional sound generation using neural discrete time-frequency representation learning,''
\newblock in {\em 2021 IEEE 31st International Workshop on Machine Learning for Signal Processing (MLSP)}. IEEE, 2021, pp. 1--6.

\bibitem{ghosal2023text}
Deepanway Ghosal, Navonil Majumder, Ambuj Mehrish, and Soujanya Poria,
\newblock ``Text-to-audio generation using instruction-tuned llm and latent diffusion model,''
\newblock {\em arXiv preprint arXiv:2304.13731}, 2023.

\bibitem{choi2023hyu}
Won-Gook Choi and Joon-Hyuk Chang,
\newblock ``Hyu submission for the dcase 2023 task 7: Diffusion probabilistic model with adversarial training for foley sound synthesis,''
\newblock Tech. {R}ep., Tech. Rep., June, 2023.

\bibitem{vaswani2017attention}
Ashish Vaswani, Noam Shazeer, Niki Parmar, Jakob Uszkoreit, Llion Jones, Aidan~N Gomez, {\L}ukasz Kaiser, and Illia Polosukhin,
\newblock ``Attention is all you need,''
\newblock {\em Advances in neural information processing systems}, vol. 30, 2017.

\bibitem{hang2023efficient}
Tiankai Hang, Shuyang Gu, Chen Li, Jianmin Bao, Dong Chen, Han Hu, Xin Geng, and Baining Guo,
\newblock ``Efficient diffusion training via min-snr weighting strategy,''
\newblock {\em arXiv preprint arXiv:2303.09556}, 2023.

\bibitem{kong2020hifi}
Jungil Kong, Jaehyeon Kim, and Jaekyoung Bae,
\newblock ``Hifi-gan: Generative adversarial networks for efficient and high fidelity speech synthesis,''
\newblock {\em Advances in Neural Information Processing Systems}, vol. 33, pp. 17022--17033, 2020.

\bibitem{ho2022classifier}
Jonathan Ho and Tim Salimans,
\newblock ``Classifier-free diffusion guidance,''
\newblock {\em arXiv preprint arXiv:2207.12598}, 2022.

\bibitem{nichol2021glide}
Alex Nichol, Prafulla Dhariwal, Aditya Ramesh, Pranav Shyam, Pamela Mishkin, Bob McGrew, Ilya Sutskever, and Mark Chen,
\newblock ``Glide: Towards photorealistic image generation and editing with text-guided diffusion models,''
\newblock {\em arXiv preprint arXiv:2112.10741}, 2021.

\bibitem{kilgour2019frechet}
Kevin Kilgour, Mauricio Zuluaga, Dominik Roblek, and Matthew Sharifi,
\newblock ``Fr{\'e}chet audio distance: A reference-free metric for evaluating music enhancement algorithms.,''
\newblock in {\em INTERSPEECH}, 2019, pp. 2350--2354.

\bibitem{chen2022beats}
Sanyuan Chen, Yu~Wu, Chengyi Wang, Shujie Liu, Daniel Tompkins, Zhuo Chen, and Furu Wei,
\newblock ``Beats: Audio pre-training with acoustic tokenizers,''
\newblock {\em arXiv preprint arXiv:2212.09058}, 2022.

\end{thebibliography}

\end{document}